\documentclass[12pt]{iopart}
\usepackage{graphicx}
\def\pmb#1{\setbox0=\hbox{$#1$}%
  \kern-.025em\copy0\kern-\wd0
  \kern.05em\copy0\kern-\wd0
  \kern-.025em\raise.0433em\box0}

\def\alt{\mathrel{\hbox{\rlap{\hbox{\lower4pt\hbox{$\sim$}}}\hbox{$<$}}}}
\begin{document}
\title{An electromagnetic analog of gravitational wave memory}

\author{Lydia Bieri}
\address{Dept. of Mathematics, University of Michigan, Ann Arbor, MI 48109-1120, USA}
\ead{lbieri@umich.edu}
\author{David Garfinkle}
\address{Dept. of Physics, Oakland University,
Rochester, MI 48309, USA}
\address{and Michigan Center for Theoretical Physics, Randall Laboratory of Physics, University of Michigan, Ann Arbor, MI 48109-1120, USA}
\ead{garfinkl@oakland.edu}


\date{\today}

\begin{abstract}
We present an electromagnetic analog of gravitational wave memory. That is, we consider what change has occurred to a detector
of electromagnetic radiation after the wave has passed.  Rather than a distortion in the detector, as occurs in 
the gravitational wave case, we find a residual velocity (a ``kick'') to the charges in the detector.  In analogy
with the two types of gravitational wave memory (``ordinary'' and ``nonlinear'') we find two types of electromagnetic
kick.

\end{abstract}


\maketitle

\section{Introduction}
\indent
Detectors of electromagnetic radiation absorb energy from the electromagnetic wave.  Since the 
flux of energy in the wave goes as $r^{-2}$ it follows that the sensitivity of the detector to a
source of a given strength falls off like $r^{-2}$.  In contrast, the sensitivity of gravitational 
wave detectors falls off like $r^{-1}$.  However, this does not have to do with some special
property of gravitational waves.  It is simply that the gravitational wave detector works not by measuring power
absorbed from the wave but rather by following the motion induced in the detector by the wave.  In principle
one could build a detector of electromagnetic waves with this same feature: simply take a charge and follow its 
motion as the wave train passes. (In practice one does not do this because such a detector would be overwhelmed by
multiple sources of electromagnetic noise).  In analogy with Christodoulou memory,\cite{christodoulou} 
one can ask what permanent change is 
induced in the charge's motion after the wave train has passed.  In the gravitational case, the wave produces
a relative acceleration between the parts of the detector.  Integrating that relative acceleration over all time, 
one might expect a residual relative velocity (a ``kick'') after the wave has passed.  However, it turns out that
such a kick is forbidden.  Nonetheless, at intermediate times there is a 
relative velocity, and one can integrate that relative velocity over all time to obtain the residual relative displacement of the parts of the detector after the wave has passed.  This residual relative displacement is the gravitational wave memory.  In the electromagnetic case, as we will see, nothing forbids a residual velocity, so in analogy to 
gravitational wave memory there is an electromagnetic wave kick.  In the next two sections we will calculate this kick,
first for the slow motion case and then in the general case.

\section{The slow motion case}

We begin by recalling some facts about gravitational wave memory in the case of weak fields and slow motion.\cite{MTW}  
For two objects in free fall with initial separation $D$ in the $i$ direction, their relative acceleration in the $j$
direction is 
\begin{equation}
{\frac {{d^2} \Delta {x^j}} {d {t^2}}} = - D {R_{titj}}
\end{equation}
However, for weak field and slow motion we have that at a large distance $r$ from the source
\begin{equation}
{R_{titj}} = {\frac {-1} r} P \left [ {\frac {{d^4}{Q_{ij}}} {d {t^4}}} \right ]
\end{equation} 
Here $Q_{ij}$ is the quadrupole moment of the source given by
\begin{equation}
{Q^{ij}} = \int {d^3} x {T^{00}} {x^i}{x^j}
\end{equation}
and $P[]$ denotes ``projected 
to be orthogonal to the radial direction and trace free.''  Any kick would then be proportional
to the difference between ${d^3}{Q_{ij}}/d{t^3}$ at large positive time and large negative time.  However, for 
large positive or negative time the only allowed motion is that of widely separated objects moving at constant 
velocity.  It then follows that for such a system
\begin{equation}
{\frac {d^2} {d {t^2}}} {Q^{ij}} = {\sum _k} {m_k}{v _k ^i}{v_k ^j}
\label{d2q}
\end{equation}  
where the sum is over all objects, where the kth object has mass $m_k$ and velocity ${\vec v}_k$.  It follows
from eqn. (\ref{d2q}) that at large positive and negative times ${d^3}{Q_{ij}}/d{t^3}$ vanishes, and therefore
there is no kick.  However, there is a residual separation given by 
\begin{equation}
\Delta {x^j} = {\frac D r} P \left [ {\sum _k} {m_k}{v _k ^i}{v_k ^j}(t=\infty) 
- {\sum _k} {m_k}{v _k ^i}{v_k ^j} (t=-\infty)\right ]
\end{equation}
This is the gravitational wave memory in the weak field, slow motion regime.  It is also what Christodoulou refers
to as the ``linear'' contribution to
gravitational wave memory, in contrast to the ``nonlinear'' memory treated in \cite{christodoulou}.

Now, consider the corresponding analysis for the motion of a charge in the presence of an electromagnetic wave.  For a
charge $q$ with mass $m$ the equation of motion is
\begin{equation}
m {\frac {{d^2} {\vec x}} {d{t^2}}} = q {\vec E}
\end{equation}
It follows that once the wave has passed the charge has received a kick given by
\begin{equation}
\Delta {\vec v} = {\frac q m} {\int _{-\infty} ^\infty} {\vec E} d t
\label{kick}
\end{equation}
However in the slow motion limit and far from the source we have\cite{LL} 
\begin{equation}
{\vec E} = {\frac 1 r} P \left [ {\frac {{d^2} {\vec p}} {d{t^2}}} \right ]
\label{Eslow}
\end{equation}
where $\vec p$ is the dipole moment of the source and $P[]$ denotes ``projected orthogonal to
the radial direction.''  It then follows that the kick is given by 
\begin{equation}
\Delta {\vec v} = {\frac q {mr}} P \left [ {\frac d {dt}}{\vec p}(t=\infty) - 
{\frac d {dt}}{\vec p}(t=-\infty) \right ]
\end{equation}
In analogy to the case of gravitational waves, we consider only systems which at large positive and negative times consist of widely separated charges each moving at 
constant velocity.  It then follows that for such a system
\begin{equation}
{\frac d {dt}}{\vec p} = {\sum _k} {q_k}{{\vec v} _k}
\end{equation}
where the sum is over all objects, where the kth object has charge $q_k$ and velocity ${\vec v}_k$.  The kick 
is therefore given by
\begin{equation}
\Delta {\vec v} = {\frac q {mr}} P \left [ {\sum _k} {q_k}{{\vec v} _k} (t=\infty) - 
 {\sum _k} {q_k}{{\vec v} _k} (t=-\infty)\right ]
\end{equation}

\section{Electromagnetic memory}

In the general case, the kick is still given by eqn. (\ref{kick}), but we can no longer use the 
far field slow motion expression 
(eqn. (\ref{Eslow})) for the
electric field.  Instead, we will need to analyze the general behavior of 
the electromagnetic field far from the source.  Recall that
the electric and magnetic fields $E_a$ and $B_a$ satisfy Maxwell's equations.
\begin{eqnarray}
{\partial _a} {E^a} = 4 \pi \rho
\\
{\partial _a}{B^a} = 0
\\
{\partial _t}{B_a} + {\epsilon _{abc}}{\partial ^b}{E^c} = 0
\\
{\partial _t}{E_a} - {\epsilon _{abc}}{\partial ^b}{B^c} = - 4 \pi {j_a}
\end{eqnarray}
where $\rho$ is the charge density and $j_a$ is the current density.  

We begin by expressing Maxwell's equations in spherical coordinates.  We denote the spherical 
coordinate indicies by $r$ if they are in the radial direction and by capital latin letters if they are in the
two-sphere direction. Then Maxwell's equations become
\begin{eqnarray}
{\partial _r}{E_r} + 2 {r^{-1}}{E_r} + {r^{-2}}{D_A}{E^A} = 4 \pi \rho
\label{divE2}
\\
{\partial _r}{B_r} + 2 {r^{-1}}{B_r} + {r^{-2}}{D_A}{B^A} = 0
\label{divB2}
\\
{\partial _t}{B_r} + {r^{-2}}{\epsilon ^{AB}}{D_A}{{E_B}} = 0 
\\
{\partial _t}{E_r} - {r^{-2}}{\epsilon ^{AB}}{D_A}{{B_B}} = - 4 \pi {j_r} 
\\
{\partial _t} {B_A} + {{\epsilon _A}^B}({D_B}{E_r} - {\partial _r}{E_B}) = 0
\\
{\partial _t} {E_A} - {{\epsilon _A}^B}({D_B}{B_r} - {\partial _r}{B_B}) = - 4 \pi {j_A}
\end{eqnarray}
Here $D_A$ and $\epsilon _{AB}$ are respectively the derivative operator and volume element of the unit two-sphere,
and all indicies are raised and lowered with the unit two-sphere metric.

We now expand all quantities in inverse powers of $r$ with expansion coefficients 
that are functions of retarded time $u=t-r$ and the angular coordinates.  For an electromagnetic field that is smooth at
null infinity we have 
\begin{eqnarray}
{E_A} = {X_A} + \dots
\label{Xdef}
\\
{B_A} = {Y_A} + \dots
\label{Ydef}
\\
{E_r} = W {r^{-2}} + \dots
\label{Wdef}
\\
{B_r} = Z {r^{-2}} + \dots
\label{Zdef}
\\
\rho = {j_r} = {r^{-2}} L + \dots 
\label{rhoj}
\end{eqnarray}
where $\dots$ means ``terms higher order in $r^{-1}$'' and we also assume that at large $r$ the angular components of 
$j_a$ are negligible compared to the radial component.
These equations require some explanation: The Cartesian components of the electric and magnetic fields fall off as $r^{-1}$.  However,
from the relation between Cartesian and angular coordinates ({\it e.g.} $z=r \cos \theta$) it follows that the angular components
of the electric and magnetic fields behave as $r^0$.  From eqn. (\ref{divE2}-\ref{divB2}) 
it then follows that the radial components of the electric and 
magnetic fields behave as $r^{-2}$.  

As for the behavior of the current and charge density, all the charged fields that we find in nature are also massive, and this along with initial data of compact support should lead to charge and current densities that 
fall off faster than any power of $r$ at
null infinity.  As we will see later, such charge and current densities would not give any analog of the 
Christodoulou memory.  However,
there is nothing wrong in principle with considering a field that is both charged and massless and this is indeed the analog for electromagnetism of fields whose stress-energy gets out to null infinity. In order to explain equation (\ref{rhoj})
we introduce the current density four-vector $J^\mu$ given by ${J^t}=\rho$ and ${J^a}={j^a}$.  
We also introduce the advanced time
$v=t+r$.  It then follows that ${J_u}= - {1 \over 2} ({j_r}+\rho)$ and ${J_v}={1 \over 2} ({j_r}-\rho)$.  Thus the behavior
given in eqn. (\ref{rhoj}) is equivalent to 
\begin{eqnarray}
{J_u}= - {r^{-2}} L + \dots
\label{Ju}
\\
{J_v}= O(r^{-3}) 
\label{Jv}
\\ 
J_{A} = O(r^{-3}) 
\label{JA}
\end{eqnarray}
Eqns. (\ref{Ju}-\ref{JA}) do not simply follow from conservation of $J^\mu$ but instead posit certain properties of $J^\mu$.
To understand the nature of these properties, note that
$-{J_u}$ is the component of the current density four-vector in the outgoing null direction and $-{J_v}$ is the 
component in the ingoing null direction.  Thus, eqns. (\ref{Ju}-\ref{JA}) amount to the statement that charge that gets out to null infinity 
does so by moving in the outgoing null direction.  Furthermore, the factor of $r^{-2}$ 
in eqn. (\ref{Ju}) is the statement that 
the rate of charge radiated per unit solid angle is finite.  
An excellent example of charges that behave in precisely this way is provided by the Maxwell-Klein-Gordon (MKG) equations for a charged, massless scalar field. The  MKG system including the behavior of its current density four-vector is discussed in 
\cite{chrdlbmathpgrt} example 3 starting on p. 99. 
Another example of charges that behave in the way we assume is that of a charged null dust.  Here, the behavior of the current density can be derived using essentially the method of \cite{lydiaandme*} (which treats the stress-energy of a null fluid), but applying that method to the charge current density of a charged null dust. 

Now putting the above expansion of the fields into the above decomposition of the Maxwell equations we obtain the following:
\begin{eqnarray}
- {\partial _u} W + {D_A}{X^A} = 4 \pi L
\\
- {\partial _u} Z + {D_A}{Y^A} = 0
\\
{\partial _u} Z + {\epsilon ^{AB}}{D_A}{X_B} = 0
\\
{\partial _u} W - {\epsilon ^{AB}}{D_A}{Y_B} = - 4 \pi L
\\
{\partial _u}{Y_A} + {{\epsilon _A}^B}{\partial _u}{X_B} = 0
\\
{\partial _u}{X_A} - {{\epsilon _A}^B}{\partial _u}{Y_B} = 0
\end{eqnarray}

The last of these equations is solved by 
\begin{equation}
{Y_A} = - {{\epsilon _A}^B}{X_B}
\end{equation}
Now plugging this result in the other equations, we find that some of the equations yield identical results and the
full set of independent equations becomes
\begin{eqnarray}
- {\partial _u} W + {D_A}{X^A} = 4 \pi L
\\
{\partial _u} Z + {\epsilon ^{AB}}{D_A}{X_B} = 0
\end{eqnarray}
Define the quantity $S_A$ by 
\begin{equation}
{S_A} = {\int _{-\infty} ^\infty} {X_A} d u
\end{equation}
Then it follows that $S_A$ satisfies the equations
\begin{eqnarray}
{D_A}{S^A} = (W(\infty) - W(- \infty)) + 4 \pi F 
\label{divS}
\\
{\epsilon ^{AB}}{D_A}{S_B} = Z(-\infty) - Z(\infty)
\label{curlS}
\end{eqnarray}
where the quantity $F$ is defined by
\begin{equation}
F = {\int _{-\infty} ^\infty} L d u 
\end{equation}
However, it also follows from the definition of $S_A$ that the kick points in the direction of $S^A$ and 
has a magnitude of 
\begin{equation}
\Delta v = {q \over {mr}} |{S^A}|
\label{kick2}
\end{equation}
(Here the magnitude of $S^A$ is calculated using the unit two-sphere metric).  
All that remains in our calculation is to show how $S^A$ depends on the quantity $F$.  Note that $F$ is the
amount of charge radiated away to infinity per unit solid angle.

In analogy to the case of gravitational waves, we consider only systems which at large positive and negative times consist 
of widely separated charges each moving at constant velocity.  For a single charge moving at constant velocity, the
$r^{-2}$ piece of $B_r$ vanishes.\cite{LL} It then follows from superposition that the same is true for a collection of
such charges, and thus it follows that both $Z(-\infty)$ and $Z(\infty)$ vanish.  it then follows from eqn. (\ref{curlS})
that there is a scalar $\Phi$ such that
\begin{equation}
{S_A}={D_A}\Phi
\end{equation} 
It then follows from eqn. (\ref{divS}) that
\begin{equation}
{D_A}{D^A}\Phi = (W(\infty) - W(- \infty)) + 4 \pi F
\label{laplacephi}
\end{equation}
Note that it is required for the consistency of this equation that the right hand side integrated over all solid angle vanishes.
In physical terms, the reason that this consistency condition is satisfied is the following: it follows from eqn. (\ref{Wdef})
that the integral over all solid angle of $W$ is $4 \pi$ times the charge enclosed.  
It then follows that the integral over all solid angle of 
$W(-\infty) - W(\infty)$ is $4 \pi$ times the amount of charge lost by being radiated to null infinity.  
But since $F$ is the charge radiated per unit solid angle, the integral over all solid angle of $4 \pi F$ is also
$4 \pi$ times the lost charge.  For any quantity on the two-sphere, we will adopt the notation that a subscript $[0]$
means the average value of that quantity.  It follows from eqn. (\ref{laplacephi}) that $\Phi$ consists of two
pieces $\Phi ={\Phi_1} +{\Phi_2}$ satisfying the following equations:
\begin{eqnarray}
{D_A}{D^A}{\Phi _1} = (W(\infty)-W(-\infty)) - {{(W(\infty)-W(-\infty))}_{[0]}}
\label{laplacephi1}
\\
{D_A}{D^A}{\Phi _2}= 4 \pi (F-{F_{[0]}}) 
\label{laplacephi2}
\end{eqnarray}  
and that ${S_A}={S_{1A}}+{S_{2A}}$ with ${S_{1A}}={D_A}{\Phi_1}$ and correspondingly for $S_{2A}$.  
In analogy with the treatment of \cite{christodoulou} we will call the kick due to $\Phi_1$ the ``ordinary'' kick and 
that due to $\Phi_2$ the ``null'' kick.  Note that $W$ at early and late times is determined by the characterization as 
a collection of charges $q_k$ with constant velocity ${\vec v}_k$.  It then follows from the treatment of \cite{LL} that  
\begin{equation}
W = {\sum _k} {q_k} (1 - {v_k ^2}) {{\left ( 1 - {\hat r}\cdot {{\vec v}_k} \right ) }^{-2}}
\end{equation}  

To find explicitly the dependence of the null kick quantity $S_{2A}$ on $F$, we will expand in spherical harmonics.  
(Note that exactly the same method yields the dependence of the ordinary kick quantity $S_{1A}$ on $W(\infty)-W(-\infty)$). 
In analogy with the method 
of\cite{christodoulou} we also provide a Green's function method in the appendix.  Expanding $F -{F_{[0]}}$ and ${\Phi_2}$ 
we have
\begin{eqnarray}
F -{F_{[0]}} = {\sum _{\ell >0}} {a_{\ell m}} {Y_{\ell m}}
\\
\Phi = {\sum _{\ell >0}} {b_{\ell m}} {Y_{\ell m}}
\end{eqnarray}
Note that it follows from the usual orthogonality of spherical harmonics that the expansion coefficients $a_{\ell m}$ are given by
\begin{equation}
{a_{\ell m}} = {\int _0 ^{2\pi}} d\phi \; {\int _0 ^\pi} \sin \theta d \theta \; {Y^* _{\ell m}}(\theta,\phi) 
(F(\theta, \phi)-{F_{[0]}})
\label{asolve}
\end{equation}
However, it follows from eqn. (\ref{laplacephi2}) that
\begin{equation} 
{b_{\ell m}} = {\frac {- 4 \pi} {\ell (\ell +1)}} {a_{\ell m}}
\end{equation}
Thus we have
\begin{equation}
{S_{2A}} =  {\sum _{\ell >0}} {\frac {- 4 \pi} {\ell (\ell +1)}} {a_{\ell m}} {D_A} {Y_{\ell m}}
\end{equation}
with $a_{\ell m}$ given by eqn. (\ref{asolve}).  Thus we have found the null part of the kick explicitly in terms of the charge radiated to infinity per unit solid angle.

\section{Conclusions}

Our results on electromagnetic kicks are very closely analogous to gravitational wave memory.  Since electromagnetism is much simpler than 
general relativity, we hope that these results can lead to greater understanding of gravitational wave memory.  In particular, it is sometimes
argued \cite{thorne,will} that there is really only one kind of gravitational wave memory: {\it i.e.} that the Christodoulou nonlinear memory is
merely the ordinary gravitational wave memory due to the energy of gravitons.  
This view however requires an artificial splitting of the vacuum Einstein equation into a linear part and a nonlinear part.
Moreover, even in the simpler case of electromagnetism we obtain two different kinds of kick: one due to the difference between the
early and late time values of the radial component of the electric field, and the other due to 
the charge radiated to infinity.  What seems crucial in our case is that the ``ordinary'' kick is due to charges that do not reach null infinity
and that the ``null'' kick is due to charges that do reach null infinity.  

It is sometimes emphasized \cite{christodoulou,lydiaetal} that gravitational wave memory is an inherently nonlinear effect that cannot be 
understood within the linearized theory.  This is certainly true for the case of memory caused by the energy of gravitational fields or the 
energy of electromagnetic fields, since in each case the energy is second order in the field.  However, we have found an analog of the 
``nonlinear'' memory even in the case of a linear theory, electromagnetism.  
On the other hand, the coupled Maxwell-Klein-Gordon system is nonlinear. 
Again, this leads us to believe that the crucial distinction is 
between an ``ordinary'' effect due to sources that do not reach null infinity, and a ``null'' effect due to sources that do.
We expect that the same distinction can be made in the gravitational 
wave memory case.  
In our forthcoming article \cite{lydiaandme} we treat ``ambiguities" in general relativity occurring from what some people may call ``linear" and ``nonlinear" features of gravitational radiation.  
We are currently applying the techniques we have used in this paper
for the electromagnetic field to the Weyl tensor of linearized gravity in the hope that this will shed further light on the nature of gravitational wave memory.\cite{lydiaandme}  

\ack
We thank Demetrios Christodoulou for discussions and valuable comments. 
DG was supported by NSF grants PHY-0855532 and PHY-1205202 to Oakland University. 
LB was supported by NSF grants DMS-1253149 and DMS-0904760 to The University of Michigan.

\appendix
\section*{Appendix}
\setcounter{section}{1}

In analogy with the treatment of \cite{christodoulou} we would like to have a Green's function method to find the dependence of the 
kick on its sources.  Recall that we want to solve
\begin{equation}
{D_A}{D^A} {\Phi _2} = 4 \pi (F - {F_{[0]}})
\label{poissonS2}
\end{equation}
Further recall that a point on the two-sphere can be represented as a unit vector $\xi$ in three dimensional Euclidean space
and that a point in Euclidean space is given by the radial distance $r$ and vector $\xi$. 
Now let $\tilde \phi$ satisfy Poisson's equation on Euclidean space
\begin{equation}
{\nabla ^2} {\tilde \phi} = - 4 \pi {\tilde \rho}
\label{poisson1}
\end{equation}
Writing Poisson's equation in spherical coordinates we obtain
\begin{equation}
{\frac \partial {\partial r}} \left ( {r^2} {\frac {\partial {\tilde \phi}} {\partial r}}\right ) + {D_A}{D^A} {\tilde \phi}
= - 4 \pi {r^2} {\tilde \rho}
\label{poisson2}
\end{equation}
Now choose the source $\tilde \rho$ to be
\begin{equation}
{\tilde \rho}(r,\xi) = ({F_{[0]}} - F(\xi)) \delta (r-1)
\label{rhoshell}
\end{equation}
and integrate eqn. (\ref{poisson2}) from $r=0$ to $r=R$ where $R$ is a large number that we will eventually take to infinity.  Then because
the charge density of eqn. (\ref{rhoshell}) has zero total charge, it follows that the first term on the left hand side of 
eqn. (\ref{poisson2}) vanishes when integrated from $0$ to $R$ in the limit $R \to \infty$.  We therefore obtain 
\begin{equation}
{D_A}{D^A} {\int _0 ^\infty } \; dr \; {\tilde \phi}(r,\xi) = 4 \pi (F (\xi) - {F_{[0]}})
\end{equation}
Comparing this equation to eqn. (\ref{poissonS2}) it follows that  
\begin{equation}
{\Phi_2}(\xi)={\int _0 ^\infty} \; dr \; {\tilde \phi}(r,\xi)
\label{solvePhi2}
\end{equation}
However, we know that the solution of Poisson's equation is 
\begin{equation}
{\tilde \phi}({\vec x}) = \int {d^3} {x'} {\frac {{\tilde \rho}({{\vec x}'})} {|{\vec x} - {{\vec x}'}|}}
\end{equation}
Switching to spherical coordinates and using eqn. (\ref{rhoshell}) we obtain
\begin{eqnarray}
{\tilde \phi}(r,\xi) = {\int _0 ^\infty} {{r'}^2} d {r'} \int d \mu ({\xi '}) {\frac 1 {|r \xi - {r'}{\xi '} |}}
({F_{[0]}} - F({\xi '})) \delta ({r'}-1)
\nonumber
\\
= \int d \mu ({\xi '}) {\frac 1 {|r \xi - {\xi '} |}}
({F_{[0]}} - F({\xi '}))
\end{eqnarray}
Here $d \mu ({\xi '})$ is the volume measure on the unit two-sphere.  Now applying eqn. (\ref{solvePhi2}) we obtain
\begin{equation}
{\Phi_2}(\xi)= \int d \mu ({\xi '}) (F({\xi '}) - {F_{[0]}}) \ln ( 1 - <\xi,{\xi '}>)
\end{equation}
where $<,>$ denotes the Euclidean inner product.
Now at $\xi$ let $w$ be a unit vector tangent to the two sphere.  Then since ${S_{2A}} = {D_A}{\Phi_2}$ it follows from eqn. (\ref{kick2})
that the component of the null kick in the $w$ direction is
\begin{equation}
{\frac q {mr}} \int d \mu ({\xi '}) ({F_{[0]}} - F({\xi '})) {\frac {<w,{\xi '}>}  { 1 - <\xi,{\xi '}>}}
\end{equation}


\section*{References}
\begin{thebibliography}{}

\bibitem{christodoulou}
Christodoulou D, {\it Phys. Rev. Lett.} 1991 {\bf 67}, 1486
 
\bibitem{MTW}
Misner C., Thorne K., and Wheeler J. {\it Gravitation}

\bibitem{LL}
Landau L. and Lifschitz E. {\it Classical Theory of Fields} 

\bibitem{chrdlbmathpgrt} 
Christodoulou D,  {\it Mathematical problems of general relativity theory I, } 
{\it EMS publishing house ETH Z\"urich. } 2008

\bibitem{lydiaandme*}
Bieri L. and Garfinkle D., preprint.

\bibitem{thorne}
Thorne, K. {\it Phys. Rev. D} 1992 {\bf 45} 520
 
\bibitem{will}
Wiseman A. and Will C. {\it Phys. Rev. D} 1991 {\bf 44}, R2945

\bibitem{lydiaetal}
Bieri L., Chen, P. and Yau. S.-T., {\it Class. Quantum Grav.} 2012 {\bf 29} 21   

\bibitem{lydiaandme}
Bieri L. and Garfinkle D., in preparation.

\end{thebibliography}
\end{document}